\documentclass[baaa]{baaa}

\usepackage[pdftex]{hyperref}
\usepackage{subfigure}
\usepackage{natbib}
\usepackage{helvet,soul}
\usepackage[font=small]{caption}

\contriblanguage{1}

\contribtype{7}

\thematicarea{3}

\title{White-dwarf asteroseismology with the {\sl TESS} space telescope}

\titlerunning{White-dwarf asteroseismology with {\sl TESS}}

\author{A.H. C\'orsico\inst{1,2}}

\authorrunning{C\'orsico}

\contact{alejandrocorsico@gmail.com}

\institute{Grupo de Evoluci\'on Estelar y Pulsaciones, Facultad de Ciencias
  Astron\'omicas y Geof\'isicas, UNLP, Argentina  
\and
Instituto de Astrof\'isica de La Plata, CONICET--UNLP, Argentina
}

\resumen{A medida que las estrellas enanas blancas se enfr\'ian, 
  atraviesan una o mas etapas de inestabilidad pulsacional en modos
  $g$ (gravedad), convirti\'endose en estrellas variables multiperi\'odicas. Los
  objetos que experimentan estos estados de inestabilidad pulsacional permiten a
  los astr\'onomos estudiar sus interiores a trav\'es de las t\'ecnicas
  astrosismol\'ogicas, como si ellos pudieran ``ver'' sus regiones internas
  analizando los espectros de per\'iodos de pulsaci\'on. La astrosismolog\'ia de
  enanas blancas ha experimentado avances extraordinarios en a\~nos
  recientes gracias a las observaciones fotom\'etricas de calidad y continuidad
  sin precedentes provistas por misiones espaciales tales como
  {\sl Kepler} y {\sl TESS}. Estos avances en el monitoreo de enanas
  blancas variables han sido acompa\~nados por el desarrollo de nuevos
  modelos estelares y t\'ecnicas para modelar sus pulsaciones.
  En el presente art\'iculo,
  revisamos los hallazgos mas importantes --hasta principios
  de 2022-- acerca de estas fascinantes estrellas pulsantes alcanzados
  gracias a las observaciones ininterrumpidas de alta calidad de la
  misi\'on {\sl TESS}, a\'un en ejecuci\'on, teniendo en mente que habr\'a
  probablemente muchos nuevos resultados en el futuro inmediato derivados
  de este telescopio espacial \'unico.}

\abstract{As white-dwarf (WD) stars cool, they go through one or more
  stages of $g$(gravity)-mode pulsational instability, becoming multiperiodic
  variable stars.
  Stars passing through these instability domains allow astronomers to
  study their interiors through asteroseismological techniques, as if
  they could ``look'' at their interiors by analyzing the spectra of
  pulsation periods. WD asteroseismology has experienced
  extraordinary advances in recent years thanks to photometric observations of
  unprecedented quality delivered by space missions such
  as {\sl Kepler} and {\sl TESS}. These advances in the monitoring of
  variable WDs have been accompanied by the development of
  new stellar models and  techniques for modeling their
  pulsations.  In this article,
  we review the most outstanding findings --up to early 2022-- about
  these fascinating pulsating stars made
  possible with the high-sensitivity and continuous observations of
  the ongoing {\sl TESS} mission, bearing in mind that there will
  possibly be many more new results in the immediate future derived
  from this unique space telescope.}

\keywords{stellar evolution, white dwarf stars, stellar interiors, stellar oscillations, asteroseismology}

\begin{document}

\maketitle

\section{Introduction}\label{S_intro}

The most frequent final chapter in the life's book of the stars
consists in the white dwarf (WD) stage. Indeed, most stars in the
Universe \citep[those with initial masses below $\sim 10-11
  M_{\odot}$;][including our Sun]{2015ApJ...810...34W} will end their
lives in the form of these hot, compact and degenerate objects that
evolve basically by cooling, delivering to the interstellar
medium their reservoir of thermal energy produced long time ago,
during their prior evolution. A comprehensive review article focused on
the origin and evolution of WDs is that  of \cite{2010A&ARv..18..471A}.
WDs are extremely old objects, with
typical ages in the range $1-10$ Gyr (1 Gyr $\equiv 10^{9}$ yr), and
are found in a range of masses of $0.15 \lesssim M_{\star}/M_{\odot}
\lesssim 1.25$, with an average value of $M_{\star}/M_{\odot} \sim
0.60$ \citep{2016MNRAS.461.2100T}.  They are characterized by
planetary dimensions, with stellar radii $R_{\star} \sim 0.01
R_{\odot}$, and thus the matter inside is highly compacted, the
average densities being of the order of $\overline{\rho}  \sim 10^6$
g/cm$^3$. The equation of state inside WDs is that of a
highly-degenerate Fermi gas \citep{1939isss.book.....C}, the
hydrostatic equilibrium in their interior being provided by the
pressure of degenerate electrons counteracting gravity. In particular,
electron degeneracy is responsible for the counter-intuitive
relationship between the stellar mass and radius, according to which
the more massive the star, the smaller its radius. Also, since the
mechanical properties of a WD are described by a Fermi gas of
degenerate electrons, there exist a limit mass ---the Chandrasekhar
mass, $\sim 1.4 M_{\odot}$--- beyond which the structure of a WD
becomes unstable.

The number of identified WDs has risen enormously in the last few
years. Ground-based observations, mainly with the spectral
observations of the Sloan Digital Sky Survey
\citep[SDSS;][]{2000AJ....120.1579Y}, have increased 15 times the
number of known WDs \citep{2019MNRAS.486.2169K}.
Recently, \cite{2021MNRAS.508.3877G} presented a
catalog of $\sim 359\,000$ high-confidence WD candidates selected from
{\it Gaia} DR3.

WDs are currently being used as valuable tools to learn about the history
of our Galaxy and stellar populations. Indeed, as nearly all stars become WDs
after the exhaustion of their central H, WDs are representative
of the age distribution of the vast majority of stars. Therefore,
they provide independent chronometers for the age and
star-formation history of the Galactic disk \citep[e.g.][]{2006AJ....131..571H},
and halo  \citep[][]{1998ApJ...503..239I} through
\emph{cosmochronology}. Furthermore, WD cosmochronology can
be applied to Galactic globular clusters  \citep{2013Natur.500...51H}
and open clusters \citep[e.g.][]{2010Natur.465..194G} as well.
WDs also are extremely useful to explore the evolution of planetary systems
\citep[see, e.g.,][]{2022MNRAS.511...71H}. On the other hand, WDs
provide us information about the  past evolution of their progenitor
stars and the chemical enrichment of the interstellar medium
\citep{2001PASP..113..409F,2008ARA&A..46..157W,2010A&ARv..18..471A}

An extremely important property of WDs, observed serendipitously for the
first time by \cite{1968ApJ...153..151L}, is that they experience
$g$-mode pulsational instabilities that make them variable stars at
least in one stage of their evolution. The interior of pulsating stars
can be studied through the tools of asteroseismology  by exploiting
the information encrypted in their frequency spectra
\citep{1989nos..book.....U,2015pust.book.....C, 2021RvMP...93a5001A,
  2022arXiv220111629K}. In the case of WDs, asteroseismology has
become in recent years one of the most important tools to learn about
their origin, evolution and internal structure, extending the reach of
the traditional techniques of spectroscopy, photometry and astrometry
\citep[see][]
      {2008ARA&A..46..157W,2010A&ARv..18..471A,2017ApJ...834..136G,2019A&ARv..27....7C,2020IAUS..357...93C}.
      In order to carry out asteroseismic analyses of pulsating WDs,
      it is crucial to have available a sufficient number of pulsation
      periods.  In the beginning, this was made a reality thanks to
      the Whole Earth Telescope \citep[{\sl
          WET};][]{1990ApJ...361..309N,1991ApJ...378..326W,
        1994ApJ...430..839W}. For a decade or so, the difficult
      undertaking of detecting as many periods as possible has begun
      to be done through  space missions. Space-based observations
      meant a revolution in the area of asteroseismology for many
      classes of pulsating stars, particularly pulsating WDs.  As a
      matter of fact, during the nominal {\sl Kepler} mission
      \citep{2010Sci...327..977B} and its {\sl K2} extension
      \citep{2014PASP..126..398H}, 89 pulsating WDs were monitored,
      and the analyses of 35 of them have been published up to now
      \citep[see ][for a complete account]{2020FrASS...7...47C}. 

\begin{figure}[h!]
\begin{center}
\includegraphics[width=8.4cm]{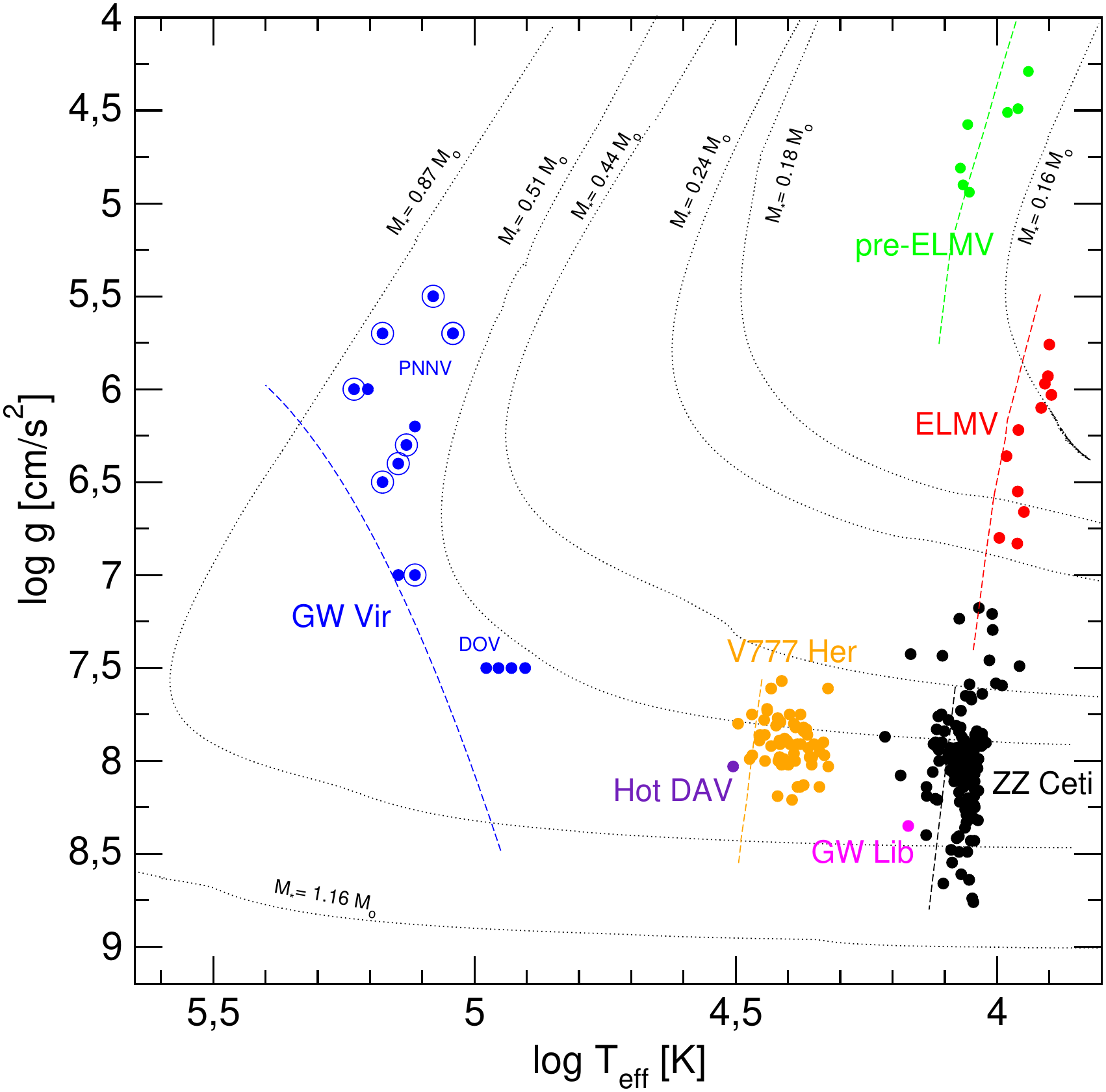}
\end{center}
\caption{Representative subsets of the different families of
  pulsating WD and pre-WD
  stars (small circles of different colors) in the $\log T_{\rm
    eff}-\log g$ plane. GW Vir stars indicated with blue circles
  surrounded by blue circumferences are PNNVs. In the case of GW Lib
  stars, only the location of the prototypical object, GW Librae, has
  been included (magenta dot). Two post-VLTP (Very Late Thermal Pulse)
  evolutionary tracks for H-deficient WDs \citep[$0.51$ and
    $0.87M_{\odot}$;][]{2006A&A...454..845M}, four evolutionary tracks
  of low-mass He-core H-rich WDs \citep[$0.16, 0.18, 0.24$,
    and $0.44 M_{\odot}$;][]{2013A&A...557A..19A}, and one
  evolutionary track for ultra-massive H-rich WDs
  \citep[$1.16 M_{\odot}$;][]{2019A&A...625A..87C} are plotted for
  reference. Dashed lines correspond to the location of the
  theoretical hot edge of the different instability strips.}
\label{fig:1}
\end{figure}

\section{The {\sl TESS} space mission}

The successor to the {\sl Kepler} spacecraft is the
{\sl NASA}'s Transiting Exoplanet Survey Satellite \citep[{\sl TESS},][]
{2015JATIS...1a4003R}. This space mission was launched on 18 April 2018 with
a planned nominal  duration of 2 years, and 15 years of extended mission, 
with the main goal of searching for exoplanets around bright stars.
Through nearly continuous stable photometry, as well as its extended sky
coverage, {\sl TESS} is making outstanding
contributions to asteroseismology. {\sl TESS} has observed 200\,000 brightest stars
in 85\% of the whole sky in 2019 and 2020 in the first part of the mission.
This space telescope performs extensive time-series photometry that allows to
discover pulsating stars, and, in particular, variable hot subdwarfs, WDs,
and pre-WDs  with mag $< 16$, with short (120\,s) cadence. Starting in July 2020,
it is now also observing in 20\,s cadence. The activities related to
compact pulsators, as WDs  and subdwarf stars, are conducted by the
{\sl TESS} Asteroseismic Science Consortium ({\sl TASC}), Compact Pulsators
Working Group (WG8). Since pulsating WDs are multiperiodic, they require
long, uninterrupted strings of data. In this sense, \cite{bradley2021}
discuss the advantages
of space satellite-based data compared to ground-based data. In the next section,
we review the most outstanding findings --up to early 2022-- on
the various classes of pulsating WDs and pre-WDs, made possible with
the {\sl TESS} space mission. 

\section{Asteroseismology of WDs and pre-WDs  with {\sl TESS}}

Pulsations in WDs and pre-WDs result in brightness variations in the
optical and also in the ultraviolet (UV) and  infrared (IR) regions of
the electromagnetic spectrum, with amplitudes between 0.001 mmag and
0.4 mag. The luminosity fluctuations (with periods in the range
$100-7000$ s) are likely induced mainly by changes in the surface
temperature ($\Delta T_{\rm eff} \lesssim 200$ K) due to spheroidal
nonradial $g$ (gravity) modes
\citep{1979ApJ...229..203M,1982ApJ...259..219R} with low harmonic
degree ($\ell= 1, 2$) and low and intermediate radial order ($k$).
Nowadays, the number of known pulsating WDs and pre-WDs amounts to
$606$ \citep{2019A&ARv..27....7C,
  2020AJ....160..252V,2021ApJ...912..125G, 2021FrASS...8..184S,
  2021ApJ...922..220L, 2021ApJ...922....2D, 2021ApJ...918L...1S,
  2022MNRAS.tmp..116R, 2022arXiv220109893V}.  They are distributed in
seven confirmed types (number of objects), namely ZZ Ceti (or DAV)
stars (494), GW Lib stars (18), V777 Her (or DBV) stars (49), hot DAV
stars (1),  GW Vir stars (24), ELMV stars (12), and pre-ELMV stars
(8).  There are one additional claimed class of pulsating WDs, the DQV
stars. The location of the different kinds of WD and pre-WDs in the
$\log T_{\rm eff}-\log g$ is depicted in Fig. \ref{fig:1}
\citep[details can be found in][]{2019A&ARv..27....7C}.

\subsection{GW Vir variables}

\begin{figure}[h!]
\begin{center}
\includegraphics[width=8.4cm]{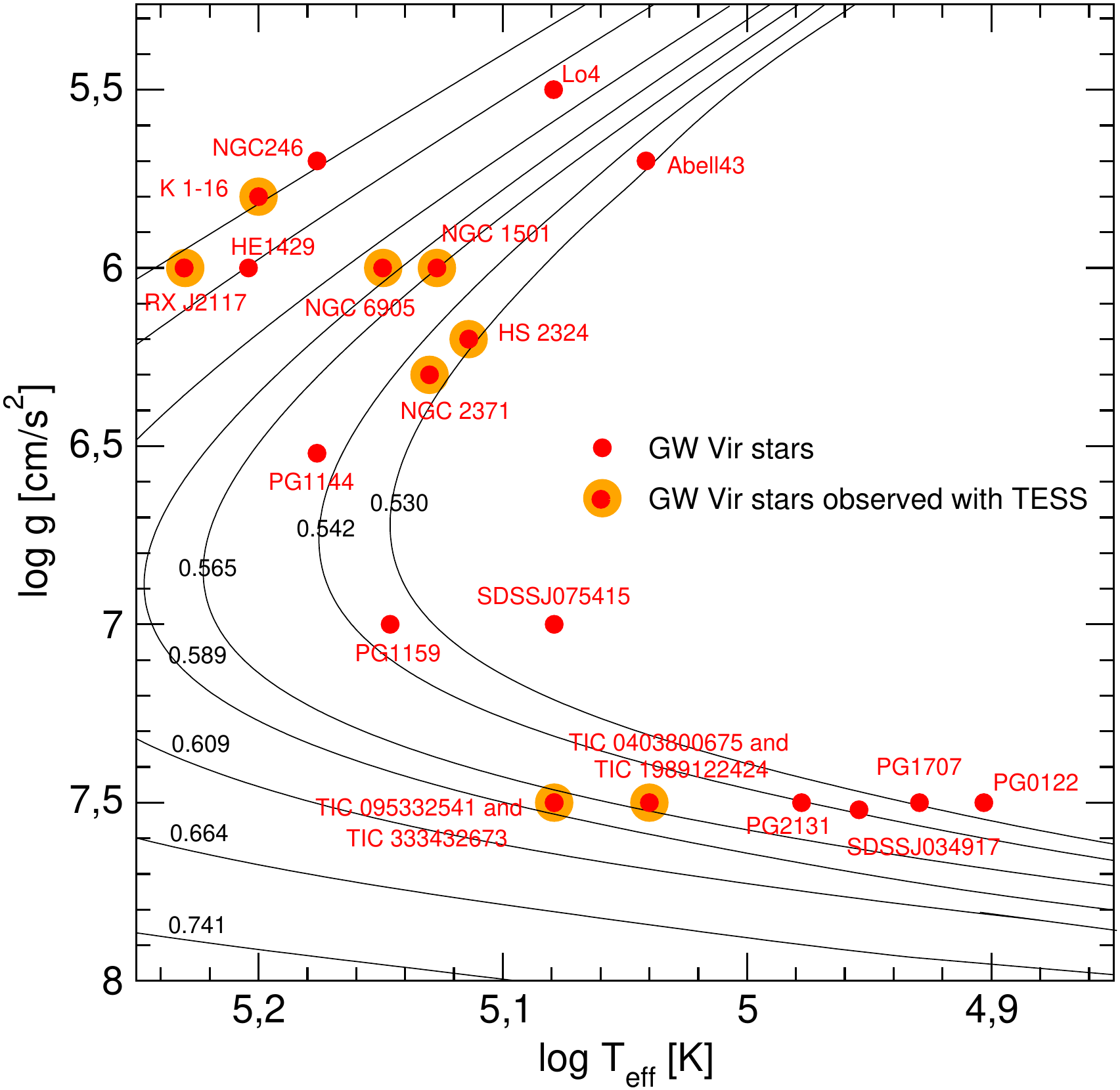}
\end{center}
\caption{Location of the known GW Vir variable stars
  in the  $\log T_{\rm eff}- \log g$
  plane depicted with red circles. Thin   solid   curves   show   the
  post-born   again   evolutionary tracks from  \cite{2006A&A...454..845M}
  for  different  stellar  masses. The location of the GW Vir stars
  observed  with {\sl TESS} are emphasized with large orange circles. 
  In particular, the stars TIC~333432673, TIC~095332541, TIC~0403800675,
  and TIC~1989122424 have been discovered with {\sl TESS} observations.}
\label{fig:2}
\end{figure}

The GW Vir variable stars are pulsating PG~1159 stars, after the prototype
of the class, PG 1159$-$035 \citep{1979ApJ...229..203M}. This is the hottest known
class of pulsating WDs and pre-WDs ($80\,000~{\rm K} \lesssim T_{\rm eff}
\lesssim 180\,000$ K and $5.5 \lesssim \log g \lesssim 7.5$), constituted 
by H-deficient, C-, O- and He-rich atmosphere WD and pre-WD stars.
This class includes stars that are still surrounded by a nebula
--the variable planetary nebula nuclei, designed as PNNVs-- and stars
that lack a nebula --called DOVs. The location of the known
GW Vir variable stars in the  $\log T_{\rm eff}- \log g$ diagram is
displayed in Fig. \ref{fig:2}.
Regarding GW Vir stars, the {\sl Kepler}
mission only observed the prototypical star of the class \citep[PG~1159$-$035;]
[]{oliveiradarosaetal2022}. At variance with this, several
GW Vir variables were scrutinized by the {\sl TESS} mission,
including four new objects discovered with this space telescope. 

In a thorough study, \cite{2021A&A...645A.117C} analyzed exhaustively
six already known GW Vir stars using observations collected by the
{\sl TESS} mission. They processed
and analyzed observations of RX~J2117+3412 (TIC~117070953), HS~2324+3944
(TIC~352444061), NGC~6905 (TIC~402913811), NGC~1501 (TIC~084306468),
NGC~2371 (TIC~446005482), and K~1$-$16 (TIC~233689607). A detailed
asteroseismological analysis of these stars was carried out on the basis
of La Plata Group's\footnote{\tt http://evolgroup.fcaglp.unlp.edu.ar/}
PG~1159 evolutionary models that take into account
the complete evolution
of the progenitor stars, this being a crucial condition to be able to
correctly assess the period spectra of this type of objects. \cite{2021A&A...645A.117C}
extracted 58 periods from the {\sl TESS} light curves of these GW Vir
stars using a standard prewhitening procedure to derive the potential
pulsation frequencies. All the oscillation frequencies that they found
are associated with $g$-mode pulsations, with periods spanning from $\sim 817$ s
to $\sim 2682$ s. The authors found a constant period spacings for all
but one star (K~1$-$16), which allowed them  to infer the stellar masses
and constrain the harmonic degree $\ell$ of the modes. Based on rotational
frequency splittings, they derived the rotation period
of RX~J2117+3412, obtaining a value in agreement with previous
determinations. An important point in this study is that to
carry out the asteroseismological analyses, the authors combined the periods
observed from the ground telescopes with those detected by {\sl TESS},
allowing them to obtain an expanded period spectrum with numerous periods
for each analyzed star. They performed period-to-period fit analyses on five of the
six analyzed stars. For four stars (RX~J2117+3412, HS~2324+3944, NGC~1501,
and NGC~2371), they were able to find an asteroseismological model with
masses in agreement with the stellar mass values inferred from the
period spacings, and are generally compatible with the spectroscopic masses.
Employing the seismological models, they  derived the seismological distance,
and compared it with the precise astrometric distance measured with {\sl Gaia},
finding good agreement in some cases. An interesting finding is that the period
spectrum of K~1$-$16 exhibits dramatic changes in frequency and amplitude, something
that made it difficult the analysis. In summary, the  findings of
\cite{2021A&A...645A.117C} confirmed the results derived for these stars on
the basis of ground-based observations. 

\cite{2021A&A...655A..27U} reported the discovery of two new GW Vir stars
with {\sl TESS} data, TIC~333432673 and TIC~095332541. Both stars are
characterized by $T_{\rm eff}= 120\,000 \pm 10\,000$ K, $\log g= 7.5 \pm 0.5$,
and $M_{\star}= 0.58_{-0.08}^{+0.16} M_{\odot}$, only differing in their
surface He/C composition. These authors presented observations from the
extended {\sl TESS} mission in both 120 s short-cadence and 20 s
ultra-short-cadence mode of these two pre-WDs showing H deficiency.
\cite{2021A&A...655A..27U} applied the tools of asteroseismology with the
aim of deriving their structural parameters and seismological distances.
The asteroseismological analysis of TIC~333432673 allowed the authors
to find a constant period spacing compatible with a stellar mass
$M_{\star} \sim 0.60-0.61 M_{\odot}$, and an asteroseismological model
for this star with a stellar mass $M_{\star}= 0.589 \pm 0.020 M_{\odot}$,
as well as a seismological distance of $d = 459_{-156}^{+188}$ pc.
For this star, there is an excellent agreement between the different
methods to infer the stellar mass, and also between the seismological
distance and that measured with {\sl Gaia} ($d_{\rm Gaia}= 389_{-5.2}^{+5.6}$ pc).
For TIC~095332541, there is a possible period spacing that suggests a
stellar mass of $M_{\star} \sim 0.55-0.57 M_{\odot}$. Unfortunately, the authors
were not  able to find an asteroseismological model for this star.

Finally, \cite{uzundagetal2022} announced the discovery of two additional
GW Vir stars, TIC~0403800675 (WD~J115727.68$-$280349.64) and
TIC~1989122424 (WD~J211738.38$-$552801.18) employing observations
collected by {\sl TESS}. These stars are characterized by
$T_{\rm eff}= 110\,000 \pm 10\,000$ K, $\log g= 7.5 \pm 0.5$,
but different He/C composition. Their {\sl TESS} light curves reveal
the presence of oscillations with periods in
a narrow range between 400 and 410\,s, which are associated with typical
$g$-modes. By performing a fit to their spectral energy distributions,
the authors found for both stars radii and luminosities of
$R_{\star}= 0.019\pm0.002\,R_\odot$ and $\log(L_{\star}/L_\odot)=1.68^{+0.15}_{-0.24}$,
respectively. Employing state-of-the-art evolutionary tracks of PG~1159 stars, they
found a stellar mass of for both stars of $0.56^{+0.15}_{-0.05} M_{\odot}$
from the $\log g-T_{\rm eff}$ diagram, and  $0.60^{+0.11}_{-0.09} M_{\odot}$
from the Hertzsprung Russell diagram. Unfortunately, due to the fact that
both stars exhibit just only two periods each, it is not possible to
perform an asteroseismological modeling, something that will have to wait
for more periods to be detected in future observations. 

\subsection{DBV variables}

\begin{figure}[h!]
\begin{center}
\includegraphics[width=8.4cm]{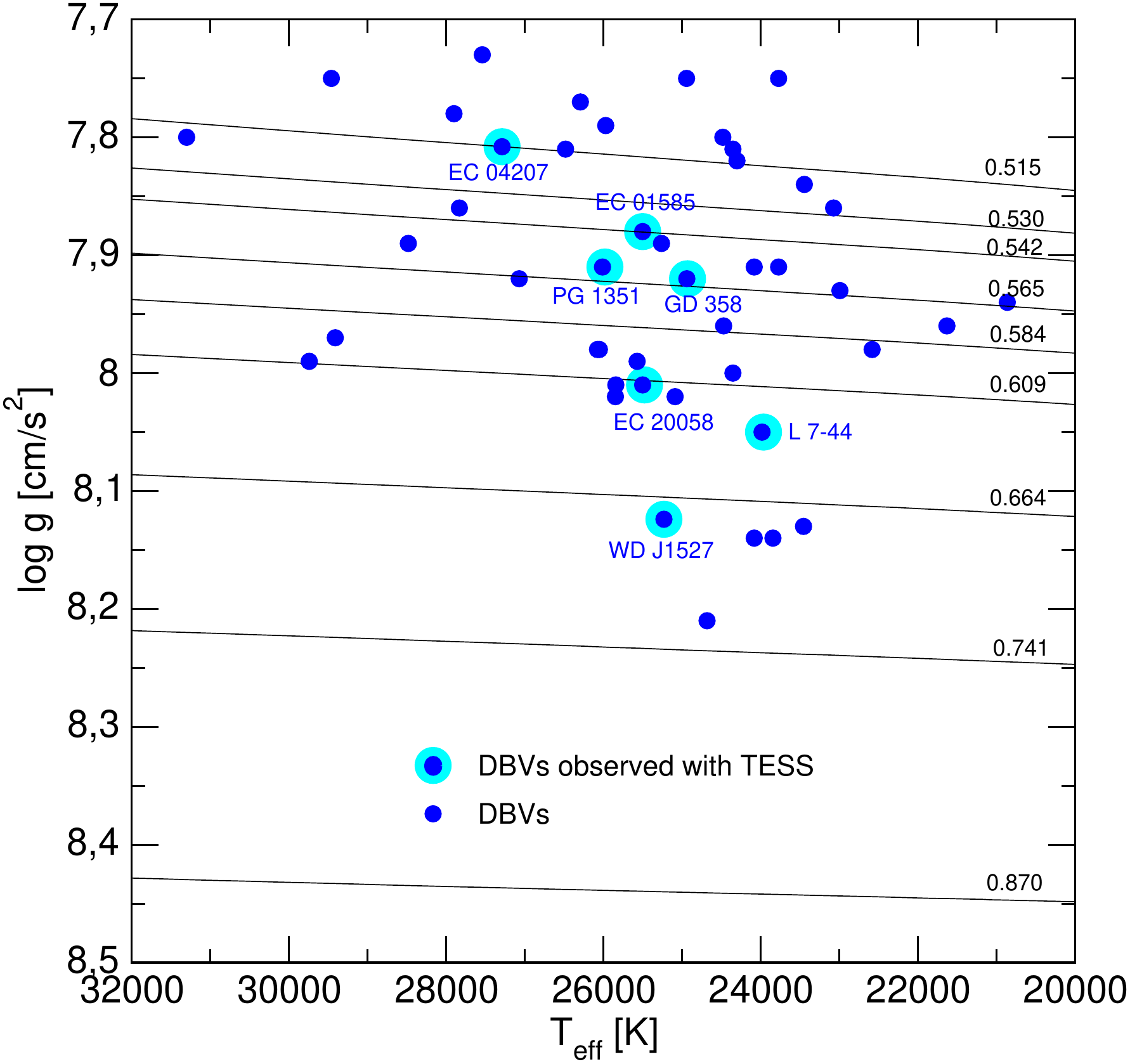}
\end{center}
\caption{Location of the known DBV WDs on the $T_{\rm eff}-\log g$ diagram,
  marked with  blue circles. Thin solid curves show the DB WD
  evolutionary tracks from \cite{2009ApJ...704.1605A} for
  different stellar masses. The location of the DBV stars observed
  with {\sl TESS} are emphasized with large cyan circles. In
  particular, the stars L~7$-$44 and WD~J1527 have been discovered
  with {\sl TESS} observations.}
\label{fig:3}
\end{figure}

The DBV or V777 Her variable stars are pulsating DB WDs, characterized by
atmospheres almost pure in He, effective temperatures in the range $22\,400~ {\rm K} \lesssim T_{\rm eff} \lesssim 32\,000$ K and surface gravities in the interval $7.5 \lesssim \log g \lesssim 8.3$. The location of the known
DBV variable stars in the  $T_{\rm eff}- \log g$ diagram 
is shown in Fig. \ref{fig:3}.

The first pulsating  WD analyzed with {\sl TESS} was the DBV star
WD0158$-$160 \citep[also called EC~01585$-$1600, G272$-$B2A, TIC~257459955;][]
{2019A&A...632A..42B}. TIC~257459955 is an already  known DBV WD with
$T_{\rm eff}= 24\,100$ K and $\log g= 7.88$ \citep{2018ApJ...857...56R}, or
alternatively, $T_{\rm eff}= 25\,500$ K and $\log g= 7.94$ \citep{2007A&A...470.1079V}.
\cite{2019A&A...632A..42B} found  9 independent periods in the range
$[245-866]$ sec, suitable for asteroseismology. The period
spacing is $\Delta \Pi= 38.1\pm 1.0$ s, which is associated to $\ell= 1$
$g$-modes. Its comparison with the average of the computed period spacings
provides an estimate of the stellar mass. This procedure points to
a stellar mass of $M_{\star}= 0.621 \pm 0.06 M_{\odot}$, or
alternatively, $M_{\star}= 0.658 \pm 0.10 M_{\odot}$,
according to the two different spectroscopic determinations of $T_{\rm
  eff}$. They  are higher than the spectroscopic estimates ($M_{\star}=
0.542-0.557 M_{\odot}$). The star also has been analyzed by means of 
period-to-period fits  by employing the fully evolutionary-model approach
of the La Plata Group,
and also the parametric approach of the Texas Group \citep{2018AJ....155..187B}.
The solution that satisfies the spectroscopic parameters and the astrometric
constraints from {\sl Gaia} is a DB WD model with $M_{\star} \sim 0.60
M_{\odot}$, $T_{\rm eff} \sim 25\,600$ K, $M_{\rm He} \sim 3 \times
10^{-2} M_{\star}$, $d \sim 67$ pc, and a rotation period of $\sim 7$
or $\sim 14$ hours.

The second DBV star studied with {\sl TESS} is the prototypical object GD~358
\citep{2021arXiv211115551C}. GD~358 (or V777 Her) has a {\it TESS} Input Catalog
(TIC) number  TIC~219074038. It is the brightest ($m_{\rm V}= 13.7$) 
and most extensively studied DBV star. Its spectroscopic surface 
parameters are $T_{\rm eff}= 24\,937\pm 1018$ K and $\log g= 7.92\pm0.05$
according to \citet{2017ApJ...848...11B} from optical data,
although the previous analysis by \cite{2012ASPC..462..171N}
and \cite{2014A&A...568A.118K} using optical and UV data gives
$T_{\rm eff}= 24\,000\pm500$ K and $\log g= 7.8\pm 0.05$ (Fig.\ref{fig:1}). 
Recently, \cite{2021RNAAS...5..249K} derived the atmospheric parameters of 
GD~358 with {\sl LAMOST} data and found $T_{\rm eff}$ = $24\,075\pm124$\,K and
$\log g = 7.827\pm0.01$\,dex. GD~358 has been extensively observed by the
{\it WET} collaboration \citep{1994ApJ...430..839W, 2000MNRAS.314..689V,
  2003A&A...401..639K,2009ApJ...693..564P}. The most recent 
and thorough analysis of this star has been carried out by \cite{2019ApJ...871...13B}, 
who through an impressive effort collected and reduced data from 34 years
of photometric observations, including  archival data from 1982 to 2006,
and 1195.2 hr of observations from 2007 to 2016.
\cite{2021arXiv211115551C} detected 26 periodicities from the {\sl TESS} light
curve of this DBV star using a standard  pre-whitening. The oscillation
frequencies are associated with nonradial $g$-mode
pulsations with periods from  $\sim 422$ s to $\sim 1087$ s. Moreover, 
they detected 8 combination frequencies between $\sim 543$ s and $\sim 295$ s.  
The authors combined these  data with a huge amount of observations from the ground
\citep{2019ApJ...871...13B} and found a constant period spacing of
$39.25\pm0.17$ s, which helped them to infer its mass
($M_{\star}= 0.588\pm0.024 M_{\sun}$) and constrain the  harmonic degree $\ell$ of the
modes. \cite{2021arXiv211115551C} performed a period-fit analysis on GD~358,
and were successful in finding an asteroseismological model with a stellar
mass $M_{\star}= 0.584^{+0.025}_{-0.019} M_{\sun}$, compatible with the stellar mass derived 
from the period spacing,  and in line with the spectroscopic mass
($M_{\star}= 0.560\pm0.028 M_{\sun}$). In agreement with previous works, 
they found that the frequency splittings vary according to the radial order of 
the modes, suggesting differential rotation. Employing the seismological model
made it possible to estimate the seismological distance
($d_{\rm seis}= 42.85\pm 0.73$ pc) of 
GD~358, which is in excellent agreement with the precise astrometric
distance measured by {\it GAIA} EDR3 ($d_{\rm GAIA}= 43.02\pm 0.04$~pc).
The authors concluded that the high-quality of data
measured with  {\it TESS}, used in combination with data taken from
ground-based observatories, provides invaluable information for conducting
asteroseismological studies of DBV  stars, analogously to what happens with
other types  of pulsating WD stars. 

Finally, an ongoing study of DBVs with {\sl TESS} is that of
\cite{uzundagetal2022b}, where a detailed asteroseismological analysis
of five DBV stars is reported. They processed and analyzed {\it TESS}
observations of the three already known DBV stars PG~1351+489 (TIC~471015205),
EC~20058$-$5234 (TIC~101622737), and EC~04207$-$4748 (TIC~153708460),
and also two new DBV pulsators WD~J1527$-$4502 (TIC~150808542) and
WD~1708$-$871 (TIC~451533898), whose variability is reported for the
first time in that paper. Similarly to what was done for the other DBVs,
in this analysis it is expected to find the stellar mass of the stars
under study through the constant period spacing  (when it exists),
and through fits of individual periods, which allow to find
asteroseismological models. The validity of these asteroseismological
models can be checked by calculating the seismological distances and comparing
them with the astrometric distances derived by {\sl Gaia}. 

\subsection{ZZ Ceti variables}

\begin{figure}[h!]
\begin{center}
\includegraphics[width=8.4cm]{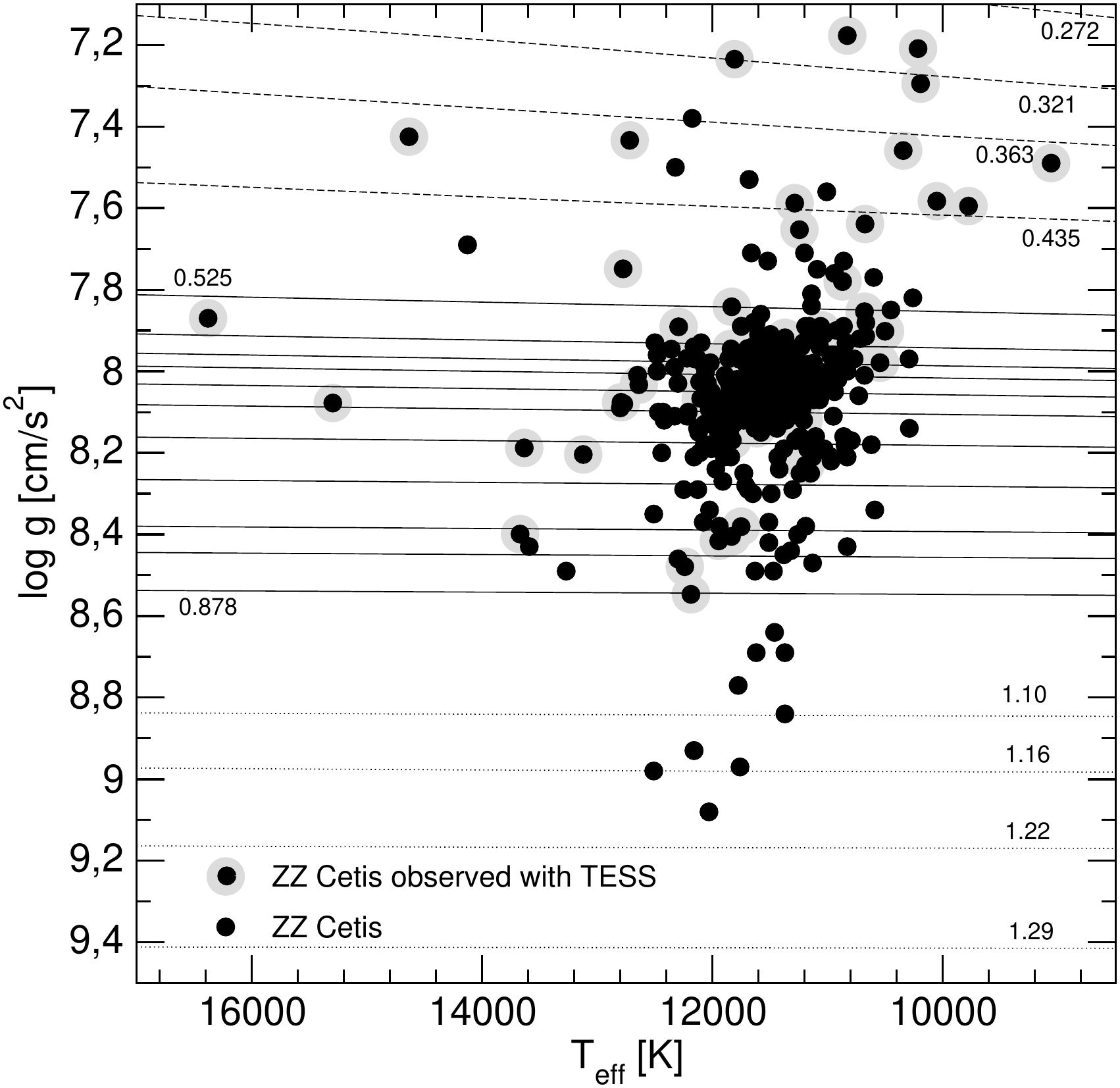}
\end{center}
\caption{Location of the known ZZ Cetis on the $T_{\rm eff}-\log g$ plane,
  depicted with black circles. Dashed curves show the low-mass He-core DA WD
  evolutionary tracks from \cite{2013A&A...557A..19A},
  solid curves depict the CO-core DA WD
  evolutionary tracks from \cite{2010ApJ...717..183R},
  and dotted curves display the ultra-massive ONe-core DA WD
  evolutionary tracks from \cite{2019A&A...625A..87C}, for different stellar masses.
  The location of the ZZ Ceti stars observed
  with {\sl TESS} are emphasized with large gray circles.}
\label{fig:4}
\end{figure}

The DAV or ZZ Ceti variable stars are pulsating DA WDs, characterized
by almost pure-H atmospheres. They are the most numerous pulsating WDs.
DAVs are located at low effective
temperatures and high gravities ($10\,400~{\rm K} \lesssim T_{\rm eff} \lesssim
12\,400$ K and $7.5 \lesssim \log g \lesssim 9.1$). It was the first class
of pulsating WDs to be detected \citep{1968ApJ...153..151L}. The location of the known
ZZ Ceti variable stars in the  $T_{\rm eff}- \log g$ diagram 
is shown in Fig. \ref{fig:4}.

The first study related to DA WDs using {\sl TESS} observations was carried out by
\cite{2020A&A...633A..20A}, although not precisely in relation to the
already known DAVs, but rather in connection to the possible existence of ``warm''
pulsating DA WDs, that is, H rich WDs with temperatures larger than those
that characterize DAVs. The motivation of \cite{2020A&A...633A..20A}
stems from the pioneering theoretical work of \cite{1982ApJ...252L..65W}, who
predicted the existence of pulsating DA WD stars with $T_{\rm eff}
\sim 19\,000$ K.  However, to date, no pulsating warm DA WD has been
detected. \cite{2020A&A...633A..20A}  re-examined the pulsational
predictions for such WDs on the basis of new evolutionary models,
 and also analyzed a sample of warm DA WDs observed by the {\sl TESS}
satellite in order to search for the possible
pulsational signals. \cite{2020A&A...633A..20A} computed WD
evolutionary sequences with extremely low H content, appropriate for the
study of warm DA WDs, and their non-adiabatic
pulsations. They  found that extended and smooth He/H
transition zones inhibit the excitation of $g$ modes due
to partial ionization of He below the H envelope, and only
in the case that the H/He transition is
assumed to be much more steep, do the models experience pulsational
instability. In this case, excited modes are found only in WDs
models with H envelopes in the range of $-14.5 \lesssim \log(M_{\rm
  H}/M_{\star}) \lesssim -10$ and at effective temperatures higher
than those typical of ZZ Ceti stars, in agreement with the previous
study by \cite{1982ApJ...252L..65W}.  \cite{2020A&A...633A..20A} found that none
of the warm DAs observed by the {\sl TESS} are pulsating. The
study suggests that the non-detection of pulsating warm DA WDs,
if WDs  with ultra-thin H envelopes do exist, could be
attributed to the presence of a smooth and extended H/He transition
zone.  This could be considered as an indirect proof that element
diffusion indeed operates in the interior of WDs. 

The second pulsational study of DA WDs carried out with {\sl TESS} observations
is that of \cite{2020A&A...638A..82B}. They presented results on 18 previously
known ZZ Ceti stars observed in 120 s cadence mode during the survey observation
of the southern ecliptic hemisphere. They compared the frequencies detected
with {\sl TESS}
with findings of previous ground-based observations, and detected possible amplitude
or phase variations during the TESS observations in some stars (EC~23487$-$2424,
BPM~30551, and MCT~0145$-$2211), something that was
not previously identified from observations observations from the ground.
Interestingly enough,
they found that HE 0532$-$5605 may be a new outbursting ZZ Ceti star \citep[see][]
{2015ApJ...809...14B}. On the other hand, they detected more than 40 pulsation
frequencies in seven ZZ Ceti stars, but did not detect any significant pulsation
frequencies in the Fourier transforms of ten observed objects, due to a combination
of their intrinsic faintness and/or crowding on the large {\sl TESS} pixels.
In particular, {\sl TESS} observations allowed the detection of  new frequencies
for five stars (EC~23487$-$2424, BPM~31594, BPM~30551, MCT~0145$-$2211,
HS~0507$+$0434B). 

The most comprehensive study of ZZ Cetis using {\sl TESS} data up to
now is that of \cite{2022MNRAS.tmp..116R}. These authors reported the
discovery of 74 new ZZ Cetis with the data obtained by {\sl TESS}
from Sectors 1 to 39, corresponding to the first 3 years of the mission.
This includes objects from the southern hemisphere (Sectors 1-13 and 27-39)
and the northern hemisphere (Sectors 14-26), observed with 120 s and
20 s cadence.  The sample of new ZZ Cetis includes 13 low-mass and one
ELM WD candidate, considering the mass determinations from fitting
{\sl Gaia} magnitudes and parallax. In addition,
\cite{2022MNRAS.tmp..116R} present follow-up time series photometry
from ground-based telescopes for 11 objects, which allowed them to
detect a larger number of periods.  For each object, they analyzed the
period spectra and performed an asteroseismological analysis,
and estimated the structure parameters ($M_{\star}, T_{\rm eff}, M_{\rm H}$) of
the sample.  They derived a mean seismological
mass of $\langle M_{\rm seis}\rangle= 0.635 \pm 0.015 M_{\odot}$, in
agreement with the mean mass using estimates from {\sl Gaia} data,
which is $\langle M_{\rm phot}\rangle= 0.631 \pm 0.040 M_{\odot}$, and
with the mean mass of previously known ZZ Cetis of $\langle M_{\star}=
0.644 \pm 0.034 M_{\odot}\rangle$. The new 74 bright ZZ Cetis
discovered by these authors increases the number of ZZ Cetis by $\sim
20 \%$, leading to 494 the total number of known
pulsating WD stars of this class. 

\subsection{Pre-ELMV and ELMV variables}

The ELMV variable stars  (Extremely Low-Mass WDs variables) have
$7800~ {\rm K} \lesssim T_{\rm eff} \lesssim 10\,000$ K and $6 \lesssim \log g
\lesssim 6.8$, pure H atmospheres, and were discovered by \cite{2012ApJ...750L..28H}.
The pre-ELMV variable stars  ($8000~ {\rm K} \lesssim T_{\rm eff} \lesssim 13\,000$
K and $4 \lesssim \log g \lesssim 5$), the probable precursors of ELMVs,
were discovered by \cite{2013Natur.498..463M}. They have H/He atmospheres. 
So far, few stars in these two categories of pulsating WD and pre-WD stars
have been studied with {\sl TESS}. 

\cite{2020ApJ...888...49W} reported the discovery of two new pre-ELMV candidates
in the {\sl TESS} eclipsing binaries TIC~149160359 and TIC~416264037.
Their lightcurves show a typical feature of EL CVn-type binaries.
The less-massive components of the two binaries are both probably
thermally bloated, pre-ELMVs. TIC~149160359 was found to pulsate in
21 independent frequencies, 4 of which are between 1139 s and 1359 s,
and are associated to pulsation modes of the pre-ELM WD component.
The Fourier amplitude spectrum of TIC~416264037 shows two pulsation
periods at 706 s and 769 s, likely corresponding to the pre-ELMV component.
Low-frequency signals (long periods) are also detected in both 
TIC~149160359 and TIC~416264037, and are probably due to the intrinsic pulsations
of their $\delta$ Sct-type primary components. 

Using observations made with the {\sl TESS} space telescope,
\cite{2021ApJ...922..220L} discovered pulsations in the ELM WD GD~278.
This star was observed by {\sl TESS} in Sector 18 at a 2-min
cadence for roughly 24\,d. The {\sl TESS} data reveal at least 19
significant periodicities between $2447-6729$\,s, one of which is
the longest pulsation period ever detected in a pulsating WD.
Previous spectroscopy found that this ELMV star is in a 4.61 hr orbit with an
unseen $> 0.4 M_{\odot}$ companion and has $T_{\rm eff}= 9230 \pm 100$ K
and $\log g= 6.627 \pm 0.056$, which corresponds to a mass of
$0.191\pm0.013 M_{\odot}$. The star exhibits clear signatures of rotational
splittings of frequencies, compatible with a stellar rotation period
of roughly 10\,hr, making GD~278 the first ELMV  with a measured rotation rate.
The spectrum of available periods does not allow finding a single
asteroseismological model. Asteroseismology reveals two main possible solutions
roughly consistent with the spectroscopic parameters of this ELM WD,
but with vastly different H-layer masses. To break this degeneracy of
solutions it would be useful to employ the stellar parallax, and thus the
astrometric distance of the star. 

\section{Conclusions}

The ongoing {\sl TESS} space mission is being very successful
in the search for extrasolar planets, but also
in the area of asteroseismology for many classes of
pulsating stars, in particular, pulsating WDs and pre-WDs.

In this work, we have briefly reviewed the results of 3 articles
focused on GW Vir stars, 3 works dedicated to DBVs, 3 studies
about ZZ Cetis, and 2 articles focused on pre-ELMVs and ELMVs.
All these works involve the discovery of new pulsating WDs and
pre-WDs and/or the follow up study of already known pulsating
stars with observations from space, for a total of 112 objects. 

Admittedly, this is just a partial account of the performance of the {\sl TESS} mission
concerning pulsating WDs and pre-WDs, covering only the first 3 years of
the mission. Undoubtedly, in the near future there will be numerous
discoveries of new pulsating WDs and pre-WDs and additional
observations of already known pulsating objects that will help to
enlarge the lists of periods available for asteroseismology.
Ongoing space missions such as {\sl TESS}, the focus of this
article, and {\sl Cheops} \citep{2018A&A...620A.203M}, as well as
future missions such as {\sl Plato} \citep{2018EPSC...12..969P}, will
participate in this enterprise.  All this wealth of knowledge will probably
help to solve some of the mysteries surrounding these ancient pulsating stars. 

%%%%%%%%%%%%%%%%%%%%%%%%%%%%%%%%%%%%%%%%%%%%%%%%%%%%%%%%%%%%%%%%%%%%%%%%%%%%%%
% Para figuras de dos columnas use \begin{figure*} ... \end{figure*}         %
%%%%%%%%%%%%%%%%%%%%%%%%%%%%%%%%%%%%%%%%%%%%%%%%%%%%%%%%%%%%%%%%%%%%%%%%%%%%%%

\begin{acknowledgement}
  I warmly thank the organizers for this excellent conference in hard times
  of pandemic. Part of this work was supported by
AGENCIA through the Programa  de  Modernizaci\'on  Tecnol\'ogica  BID  1728/OC-AR,
by  the  PIP  112-200801-00940 grant from CONICET, and  by the grant  G149  from
University  of  La  Plata. This research has made intensive use of NASA Astrophysics
Data System.
\end{acknowledgement}

%%%%%%%%%%%%%%%%%%%%%%%%%%%%%%%%%%%%%%%%%%%%%%%%%%%%%%%%%%%%%%%%%%%%%%%%%%%%%%
%  ******************* Bibliografía / Bibliography ************************  %
%                                                                            %
%  -Ver en la sección 3 "Bibliografía" para mas información.                 %
%  -Debe usarse BIBTEX.                                                      %
%  -NO MODIFIQUE las líneas de la bibliografía, salvo el nombre del archivo  %
%   BIBTEX con la lista de citas (sin la extensión .BIB).                    %
%                                                                            %
%  -BIBTEX must be used.                                                     %
%  -Please DO NOT modify the following lines, except the name of the BIBTEX  %
%  file (whithout the .BIB extension).                                       %
%%%%%%%%%%%%%%%%%%%%%%%%%%%%%%%%%%%%%%%%%%%%%%%%%%%%%%%%%%%%%%%%%%%%%%%%%%%%%% 

\bibliographystyle{baaa}
\small
\bibliography{bibliografia}
 
\end{document}